# Detecting planetary geochemical cycles on exoplanets: Atmospheric signatures and the case of $SO_2$

L Kaltenegger and D Sasselov
*Harvard University, 60 Garden Street, 02138 MA, Cambridge USA*
*Email: lkaltene@cfa.harvard.edu*

**Abstract**
We study the spectrum of a planetary atmosphere to derive detectable features in low resolution of different global geochemical cycles on exoplanets - using the sulphur cycle as our example. We derive low resolution detectable features for first generation space- and ground- based telescopes as a first step in comparative planetology. We assume that the surfaces and atmospheres of terrestrial exoplanets (Earth-like and super-Earths) will most often be dominated by a specific geochemical cycle. Here we concentrate on the sulphur cycle driven by outgassing of $SO_2$ and $H_2S$ followed by their transformation to other sulphur-bearing species which is clearly distinguishable from the carbon cycle which is driven by outgassing of $CO_2$. Due to increased volcanism, the sulphur cycle is potentially the dominant global geochemical cycle on dry super-Earths with active tectonics. We calculate planetary emission, reflection and transmission spectrum from 0.4 to 40 μm with high and low resolution to assess detectable features using current and Archean Earth models with varying $SO_2$ and $H_2S$ concentrations to explore reducing and oxidizing habitable environments on rocky planets. We find specific spectral signatures that are observable with low resolution in a planetary atmosphere with high $SO_2$ and $H_2S$ concentration. Therefore first generation space and ground based telescopes can test our understanding of geochemical cycles on rocky planets and potentially distinguish planetary environments dominated by the carbon and sulphur cycle.

**Introduction**
The study of planets orbiting nearby stars has entered an era of characterizing massive terrestrial planets (aka super-Earths) with the recent large sample of detections (Mayor et al. 2008) and the first rocky transiting system CoRoT-7b (Leger et al. 2009). Transmission and emergent spectra of terrestrial exoplanets, especially super-Earths, could be observed in the near future with the same transiting geometry technique that proved so successful in providing spectra of giant gas exoplanets recently (see e.g. Grillmair et al. 2008 Swain et al. 2008). Such spectroscopy provides molecular band strengths of multiple detected transitions (in absorption or emission) of a few abundant molecules in the planetary atmosphere. Absolute fluxes in the IR (complemented by data in the optical if possible) are also measured often at different planetary longitudes allowing crude model atmospheres to be constructed. Here we ask how such information in emergent and transmission spectra of Earths and super-Earths could be used to explore the geochemical processes operating on the surface of rocky exoplanets remotely, as a first step in comparative planetology to test our understanding of geochemical cycles on rocky planets.

Super-Earths are a diverse class of planets that differ from giant planets (Jupiter- and Neptune-like ones) in that they have a surface - a solid-to-gas or liquid-to-gas phase transition at their upper boundary similar to Earth. That surface separates a vast interior reservoir (e.g. Earth's mantle) from an atmosphere which has insignificant mass compared to the planet's mass. Like on our own planet the atmosphere is fed from the interior reservoir and its chemical balance is achieved from interactions with the interior (outgassing and burial) and with the parent star (photochemistry and loss to space). These interactions are usually described in terms of geochemical cycles. The Earth's carbon cycle is one of them which is maintained by the outgassing of $CO_2$.

In this paper we study the spectrum of a planetary atmosphere on planets where surface conditions are conducive to accumulation of ppm levels of $SO_2$ and corresponding levels of $H_2S$. Our goal is to determine if any specific spectral signatures are observable and how well future measurements could distinguish rocky Earth-like planets dominated by different geochemical cycles using the carbon and sulphur cycle as an example here.

The organization of this paper is as follows. Section 1 introduces the sulphur and carbon geochemical cycles section 2 outlines our spectroscopic and atmospheric models and presents the corresponding atmospheric gas and temperature profiles for current and Archean Earth. Section 3 presents the spectra and results of our models for high $SO_2$ and $H_2S$ concentrations and Section 4 discusses the results of our models and detectabilty of a global sulphur cycle.

## 1. Planetary geochemical cycles

Global planetary geochemical cycles are few in type because there are a limited number of abundant gases that could be cycled in any planet regardless of environmental conditions. For rocky Earths and super-Earths in the habitable zones of their stars, namely in an equilibrium temperature range that allows liquid surface water, the geochemical cycles will involve a hydrological component.

The carbon cycle e.g. on early pre-biotic Earth would be fed by the volcanic outgassing of $CO_2$ which is balanced by the burial of calcium carbonate through silicate weathering reactions that remove protons and release alkalinity to water in oceans (Walker Hays & Kasting 1981). The spectral signatures of a global carbon cycle on the emergent spectrum of a planet over geological time have been computed by Kaltenegger et al. (2007). Here we focus on an alternative global cycle that might compete for dominance in some planets - the sulphur cycle. In the sulphur cycle volcanic release of $SO_2$ and $H_2S$ would be balanced by the precipitation of sulphite minerals (e.g. calcium sulphite $CaSO_3$) and photochemical removal (Halevy, Zuber & Schrag 2007). Increased volcanic activity and outgassing of $CO_2$, $SO_2$, and $H_2S$ could exhaust the oxidant supply, allowing $SO_2$ to reach concentrations of several hundred ppm near the planet's surface (see also Halevy et al. 2007 and Kasting et al. 2009). When volcanism subsides, $SO_2$ would be rapidly removed from the atmosphere by continued photolysis and by reaction with oxidants, now supplied at a faster rate than reduced gases. On a dry planet, the lifetime of sulphur component in the atmosphere would increase (see discussion).

$SO_2$ is a greenhouse gas and the dependence of weathering rates on atmospheric $pSO_2$ creates a climate feedback similar to that involving $CO_2$: if the volcanic release of $SO_2$ increases the temperature increases because of the greenhouse effect of the higher atmospheric $pSO_2$ the increased temperature causes an acceleration of the weathering reactions and increased removal of $SO_2$ by precipitation of sulphite minerals. This leads to a balance and stabilizes surface temperatures. Halevy et al. (2007) constructed a model of the sulphur and carbon cycles for early wet Mars to explore this feedback. Generalization of their results implies that on certain exoplanets with vigorous outgassing of $CO_2$, $SO_2$ and $H_2S$ the oxidant supply could be exhausted leading to concentrations of $SO_2$ of a few to several hundred ppm and a dominant sulphur cycle.

On dry super-Earths the sulphur cycle might be the dominant global geochemical cycle, due to their higher and longer-lasting outgassing fluxes. In addition super-Earths close to their host stars that have lost a big part of their water reservoir, thus reducing precipitation, could also be dominated by the sulphur cycle. Higher volcanic and vent outgassing is expected on rocky super-Earths because of increasing plate tectonic activity with increasing planet mass for super-Earths (Valencia, O'Connell & Sasselov 2006).

## 2. Atmospheric Model

We use EXO-P a coupled one-dimensional code developed for rocky exoplanets based on the

1D climate (Kasting & Ackerman 1986, Pavlov et al 2000, Haqq-Misra et al. 2008) 1D photochemistry (Pavlov & Kasting 2002) and radiative transfer model (Traub & Stier 1978, Kaltenegger & Traub 2009) to self consistently calculate the hypothetical spectrum of rocky exoplanet dominated by the sulphur cycle. In this paper we concentrate on the detectabilty of $SO_2$ and $H_2S$ from 0.4 μm to 40 μm to inform future ground and space based mission like the Extreme Large Telescope (ELT) and the James Webb Telescope (JWST).

The high $SO_2$ concentration models presented here are not based on fully self consistent calculations because the sulphur cycle and its detailed photochemistry is not yet fully understood and our goal at this time is to simply study the effect of a radiatively significant (1-10ppm $pSO_2$) sulphur concentration on the emergent and transmission spectral signatures of a planet and compare them to a planet (current and early Earth) with a dominant carbon cycle. Levels below 0.1ppm are too low to dominate the geochemical cycle while levels above 10ppm will most likely influence the temperature profile significantly. Therefore we model concentrations between 1 and 10ppm. We do not take the $SO_2$ greenhouse effect into account in the temperature profile calculations. High $SO_2$ concentration has been discussed as a mechanism to e.g. warm early Mars (see Kasting et al. 2009, Halevy et al. in prep). Our adopted model atmosphere is current Earth as well as a self consistently calculated Archean atmosphere with an artificially increased $SO_2$ mixing ratio of 1ppm and 10ppm and the corresponding $H_2S$ abundance.

On a rocky exoplanet increased sulphur mixing ratio can be achieved by an increase in outgassing flux or a decrease in photochemical sinks and burial rates. Current Earth mixing ratio is $2.4 \; 10^{-10}$ for $SO_2$ and $5.52 \; 10^{-8}$ for $H_2S$ (Segura et al. 2003) what corresponds to a sulphur outgassing flux of $10^9$ S atoms $cm^{-2}s^{-1}$ (Bluth et al. 1993) but this sulphur is emitted in an oxidizing atmosphere. The sulphur photochemistry in an anoxic atmosphere can have two general outcomes: (a) the production of sulphuric acid $H_2SO_4$ when some oxidation can occur e.g. due to OH and O radicals produced by photolysis of $H_2O$, $CO_2$ and $SO_2$ and (b) $SO_2$ photolysis followed by disproportionation to sulphate ($S^{6+}$) e.g. $H_2SO_4$ and elemental sulphur and sulphur chains ($S_2$ $S_4$ $S_8$) both of which produce UV absorbing aerosols (Kasting et al. 1989) as well as $H_2S$ if the atmospheric $H_2$ concentration is high (Kasting et al. 2009).

Model spectra are calculated with a radiative transfer code based on the Smithsonian Astrophysical Observatory code originally developed to analyze balloon-borne far-infrared thermal emission spectra of the stratosphere and later extended to include visible reflection spectra (Traub & Stier 1976, Jucks et al. 1998, Traub & Jucks 2002, Kaltenegger et al 2007, Kaltenegger & Traub 2009). The spectral line database includes the large HITRAN 2008 compilation plus improvements from pre-release material and other sources (Rothman et al. 2004, 2009, Yung & DeMore 1999). The far wings of pressure-broadened lines can be non-Lorentzian at around 1000 times the line width and beyond therefore in some cases ($H_2O$, $CO_2$, $N_2$) we replace line-by-line calculation with measured continuum data in these regions. Rayleigh scattering is approximated by applying empirical wavelength power laws (Allen 1976, Cox 2000) that exert an appreciable effect in the visible blue wavelength range. We have not included aerosol scattering or hazes in the modelled spectrum due to the uncertain increase of aerosol quantities versus altitude in the sulphur cycle (this should be studied in a separate paper). Therefore the estimated depth of the atmospheric features can be slightly overestimated depending on the altitude the aerosols accumulate at. We assume that the light paths through the atmosphere can be approximated by incident parallel rays bent by refraction as they pass through the atmosphere and either transmitted or lost from the beam by single scattering or absorption. We model the Earth's spectrum using its spectroscopically most significant molecules $H_2O$, $O_3$, $O_2$, $CH_4$, $CO_2$, CO, $H_2S$, $SO_2$, OH, $H_2O_2$, $H_2CO$, O, $HO_2$, $CH_3Cl$, HOCl, ClO, HCl, $NO_2$, NO, $HNO_3$, $N_2O_5$, $N_2O$ and $N_2$ where $N_2$ is included for its role as a Rayleigh scattering species and broadening gas. We do line-by-line radiative transfer through the refracting layers of the atmosphere.

Clouds are represented by inserting continuum-absorbing/emitting layers at appropriate altitudes and broken clouds are represented by a weighted sum of spectra using different cloud layers. We assume that the complex cloud distribution horizontally and vertically can be approximated by parallel streams. All streams traverse the same molecular atmosphere but each stream reflects/absorbs from a different lower surface. The cloud layers are set at an adjustable height (here 1 km 6 km and 12 km modelled on present Earth). On Earth there is an overall 60% cloud coverage. The relative proportions of each stream were set to be consistent with the Earthshine data discussed earlier. We assume in our model that the fraction and nature of clouds has not changed appreciable over time. The radiative transfer models have been successfully used to fit data from ground and space based missions (see e.g. Woolf et al. 2002, Turnbull et al 2006, Kaltenegger & Traub 2009) and to model the spectrum throughout geological evolution of our own planet (Kaltenegger et al. 2007).

### *2.1 Current and Archean Earth Model*

For current Earth we use the US Standard Atmosphere 1976 spring-fall pressure-temperature profile (COESA 1976, Cox 2000).

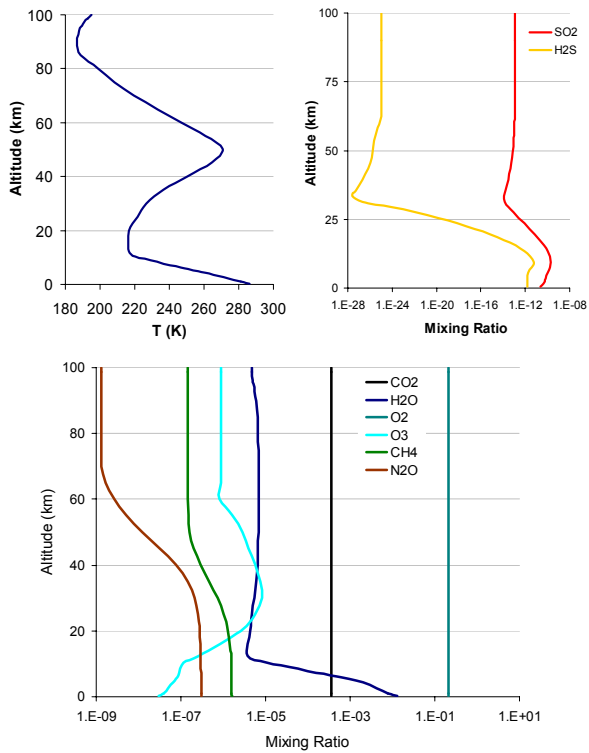

**Figure 1:** Temperature profile (top left) from US Standard Atmosphere 1976 (spring/autumn) (COESA 1976) and mixing ratios versus height for the major detectable atmospheric gases up to 100 km for (top right, bottom panel).

Figure 1 and figure 2 show the temperature profile as well as the concentration profiles of the spectrally most significant atmospheric molecules. The surface pressure is set to 1 bar. We calculate the Archean Atmosphere structure with a very low oxygen mixing ration of $10^{-7}$ ($10^{-5}$ PAL) $O_2$, $3.6 \cdot 10^{-2}$ $CO_2$ (100 PAL) and $1.7 \cdot 10^{-4}$ $CH_4$ (100 PAL). For further details on the model in an exoplanet context see Kaltenegger et al. (2007, 2009). We divide the atmosphere into 30 thin layers from 0-100 km altitude with thinner layers closer to the ground. The spectrum is calculated at very high spectral resolution with several points per line width and often thousands of points in the wings. The line shapes and widths are computed using Doppler and pressure broadening on a line-by-line basis. Scattered photons are treated as lost to absorption since we do not include multiple scattering in this code.

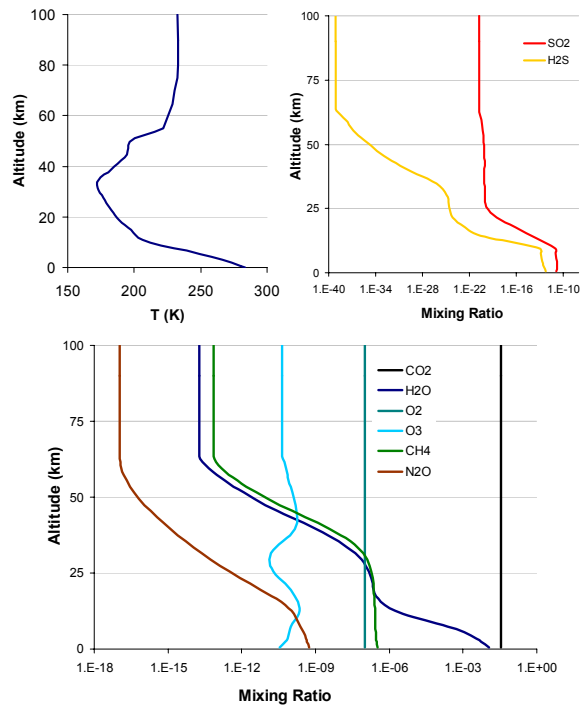

**Figure 2:** Temperature profile (top left) and mixing ratios versus height for the major detectable atmospheric gases up to 100 km for an Archean Earth atmosphere model (top right , bottom panel).

### 3: Results

We generate visible to mid-infrared spectra from 0.4 μm to 40 μm to assess detectable features of high $SO_2$ and $H_2S$ concentrations for current and

Archean Earth. To explore if features of the sulphur cycle could be detectable on rocky extrasolar planets we artificially increased the abundance of $SO_2$ to 1ppm and 10ppm and corresponding $H_2S$ abundance. The results are shown in figure 3 for emerging and in figure 4 for transmission spectra.

The fundamental vibrational frequencies of $SO_2$ and $H_2S$ shown in figure 3 and figure 4 are at 8.68 μm 19.30 μm, 7.34 μm and at 3.82 μm as well as 8.45 μm and 3.80 μm respectively sorted by their strength, placing them in the mid IR. We only show the calculations from 4 to 40micron in figure 3 to figure 6 because we concentrate on detectable features of the sulphur cycle here (for atmospheric signatures of a model of Earth through geological time in the visible to near IR see Kaltenegger et al. 2007). Additional detectable feature in the mid-IR in low resolution are the 9.6 μm $O_3$ band, the 15 μm $CO_2$ band and the 6.3 μm $H_2O$ band or its rotational band that extends from 12 μm out into the microwave region, a methane feature at 7.66 μm and three $N_2O$ features at 7.75 μm, 8.52 μm and 16.89 μm.

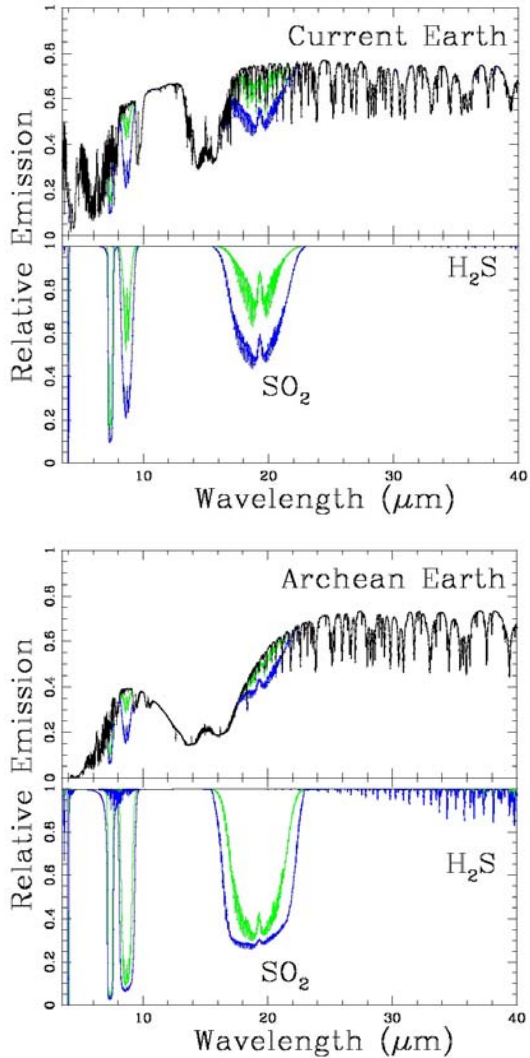

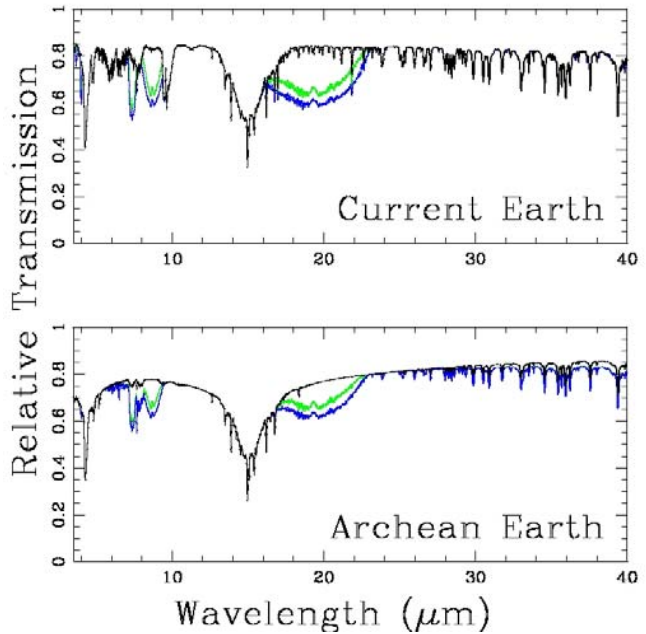

Figure 3 A calculated reflection and thermal emission spectrum shown from 4μm – 40μm of a cloud-free atmosphere for Current (top) and a model of Archean Earth (bottom) – the region from 0.4 to 4 micron shows not detectable features of the sulphur cycle in low resolution and is not shown here. The lower panel in each figure shows the $SO_2$ and $H_2S$ species 1PAL (black), 1ppm (green) and 10ppm (blue). The calculations are performed at very high resolution (0.1 wavenumbers) and subsequently smoothed for display (R = 150).

Figure 4 A calculated transmission spectrum from 4μm – 40μm for Current (top) and Archean Earth (bottom panel) shows the $SO_2$ and $H_2S$ species 1PAL (black), 1ppm (green) and 10ppm (blue). The calculations are performed at very high resolution (0.1 wavenumbers) and subsequently smoothed for display (R = 150). The region from 0.4 to 4 micron shows not detectable features of the sulphur cycle in low resolution and is not shown here.

## 4. Discussion

Here we asked if different global chemical cycles are potentially detectable in emergent and transmission spectra of rocky Earths and super-Earths. This could be used to explore the basic geochemistry of rocky exoplanets. $SO_2$ is detectable in an exoplanet's emergent and transmission spectrum. For future first generation space- and ground- based missions only strong features will be detectable. To determine which features can be detectable in very low resolution (R = 25) we show the spectra of current and Archean Earth versus a rocky exoplanet with high $SO_2$ and $H_2S$ concentration indicating a dominant sulphur cycle in emitted (figure 5) and transmitted spectra (figure 6). We show that levels of 1ppm could be remotely observed with low resolution.

### 4.1 Under what conditions could a global Sulphur cycle dominate?

Earth's current volcanic emission rates are estimated to 8 megatons of $SO_2$ per year (Kasting et al. 1989), what translates into a total surface flux of 1 $10^9$ S atoms $cm^{-2}$ $sec^{-1}$. The largest lava flow in Earth's recorded history, the 1783–1784 flood basalt eruption associated with the Laki volcanic fissure in Iceland, released 15 $km^3$ of basalt and 122 megatons of $SO_2$ into the atmosphere over a period of 8 months, with nearly half released in the first 6 days (Thordarson & Self 2003). We estimate the amount of $SO_2$ for a planet like Earth with an oxidizing atmosphere with one eruption per year as well as increased volcanism to asses the potential $SO_2$ levels. For one such eruption per year, the $SO_2$ is only about 15 times current: For a planet with high tectonic activity and such an eruption every week, the amount of $SO_2$ outgassing is about 800 times current. This still assumes only one such volcano on the planet and an oxidizing atmosphere – the second assumption is the limiting factor in the build up of $SO_2$. In addition a dry environment would prolong the lifetime of sulphur components in the atmosphere. Keeping the assumption of an oxidizing atmosphere and a water reservoir comparable to Earths, increased volcanism (e.g. 100 volcanoes) would that amount to 1500 to 80.000 times respectively and 0.15ppm to 8 ppm $SO_2$. This seems unlikely, but increased tectonic activity could increase volcanic outgassing especially on super-Earths. In our Solar System the biggest known volcano is Olympus Mons, with a volume of about 70 times that one Earth's biggest volcano Mauna Loa, assuming a cone volume. The eruption of one such volcano on a planet per year would increase the amount of $SO_2$ 1000 times to 0.1ppm. Increase volcanic activity as expected on rocky super-Earths may show an increased sulphur outgassing flux that could be achieved based on estimates of the shorter plate tectonic timescales and length of subduction boundaries for super-Earths (Valencia et al. 2007).

Note that that ratio would increase if we assume a reduced atmosphere like Archean Earth. On a planet with very active tectonic activity, volcanic outgassing of sulphur to the atmosphere and surface environment, primarily as $SO_2$ and $H_2S$, would be balanced by photochemical sinks and precipitation of sulphur-bearing minerals. The lifetime of sulphur species that today are rapidly oxidized in Earth's atmosphere is longer under reducing conditions like on Early Earth (see Archean Earth calculations). Increased volcanic activity and outgassing of $CO_2$, $SO_2$, and $H_2S$ could exhaust the oxidant supply in a reduced atmosphere, allowing $SO_2$ to reach concentrations of several hundred ppm near the planet's surface (see also Kasting et al. 2009). The simulations have been done for early Mars. The local atmospheric chemical impact of a volcanic event on the present oxidizing Martian atmosphere results in a mixing ratio of more than 100 ppm $SO_2$ in the lower 60 km of the atmosphere and a very low photolysis rate constant of $6.1 \times 10^{-18}$ $s^{-1}$ at an altitude of 10 km (Wong et al.2003).

### 4.2 Detectable Features in low resolution

In primary transmission the increase of the surface of the annulus the starlight shines through is counteracted by the increased gravity on a super-Earth leading in first approximation to a similar spectrum than an Earth-size exoplanet assuming similar mixing ratios and atmospheric profiles (Kaltenegger & Traub 2009). In addition in emergent spectra the increase in surface area makes it easier to detect super-Earths because they emit and reflect more photons than small Earth-analogs making super-Earths excellent targets for secondary eclipse measurements and direct imaging. Therefore secondary eclipse measurement and direct imaging of large super-Earths are potentially the first steps to assess whether different chemical cycles can be detected on rocky exoplanets with future missions like JWST.

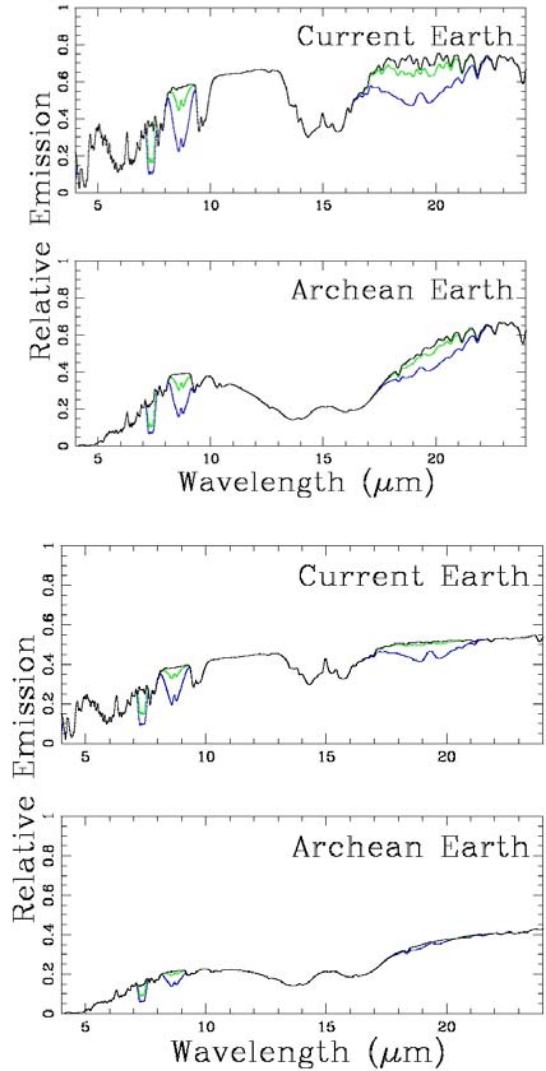

**Figure 5** Low resolution (R=25) emergent spectra of a model atmosphere without (top) and including clouds (bottom) shows the detectable features of the sulphur cycle in an emergent spectrum for secondary eclipse measurements and direct imaging with future missions like JWST: 1PAL (black), 1ppm (green) and 10ppm (blue).

Our findings that significant differences exist in the detectable features of rocky planets that are dominated by a global sulphur cycle or a carbon silicate cycle warrant a careful coupled and complete model calculation in order to determine the strength of discriminating spectral features for a variety of planetary parameters and conditions. High $SO_2$ concentrations will most probably coincide with high $CO_2$ in the atmosphere due to the relative stability of $CO_2$ and assuming a reasonable S:C in volcanic outgassing. Which cycle ends up dominating the climate and the mineral record in reality depends on many factors including the temperature on the planet, it composition and whether the surface is rocky or watery, here we only present models for rocky planets that are Earth-analogs.

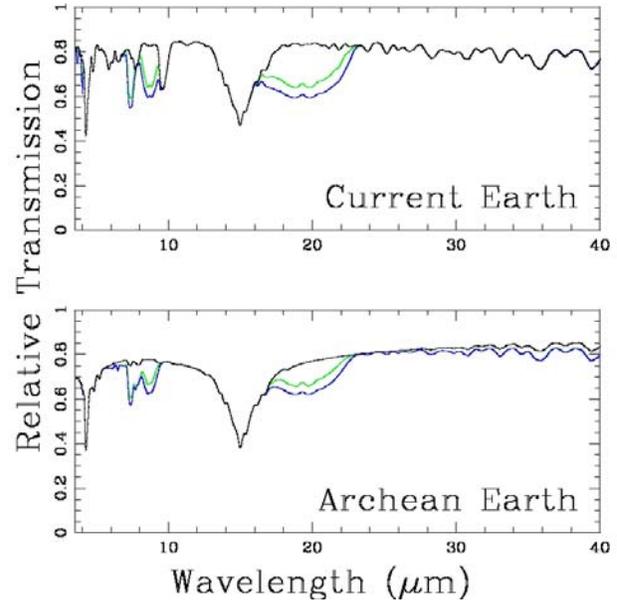

**Figure 6** Low resolution (R=25) transmission spectra of a model atmosphere that shows the detectable features in a transmission spectrum for primary eclipse measurements with future missions like JWST: 1PAL (black), 1ppm (green) and 10ppm (blue). Because the lower atmosphere is mostly opaque, low clouds do not effect the detectable feature,

**Conclusions**

We propose that the surfaces and atmospheres of terrestrial exoplanets (Earth-like and super-Earths) will most often be dominated by a specific geochemical cycle. Could such cycles be distinguished by the low-quality spectroscopy of their atmospheres we hope to obtain in the future? Here we show that in high concentrations features of $SO_2$ are detectable in low resolution in a rocky exoplanet atmosphere and we can characterize planetary environments and potentially distinguish between a carbon and a sulphur cycle through atmospheric signatures. A planet dominated by the sulphur cycle where surface conditions are conducive to accumulation of ppm levels of $SO_2$ in the atmosphere is clearly distinguishable from the carbon cycle which is driven by outgassing of $CO_2$ in our atmosphere model. More than one spectral feature of $SO_2$ becomes clearly visible in the spectrum between 0.4 and 40 μm for concentrations that exceed a few ppm (the region from 0.4 to 4 micron shows not detectable features of the sulphur cycle in low resolution and is not shown in figure 3 to figure 6). The relative strengths of broad $SO_2$, $CO_2$ and $H_2O$ features can then be used to infer the radiative dominance of the greenhouse $SO_2$ gas for the planet in question (figure 5 and figure 6) with very low resolution spectroscopy. The spectral lines of $H_2S$ are weaker and thus much more difficult to observe in this spectral range. Fortunately measurement of $H_2S$ concentrations is not critical in establishing a possible sulphur cycle on a planet.

More work on other cycles and environments is necessary to be confident that the signatures shown here map uniquely onto these cycles. Accumulation of $SO_2$ in a $CO_2$-rich atmosphere, made possible by small total photochemical sink and by saturation of the surface, could also expand the limit of the outer edge of the Habitable Zone around a star further out than the Carbon silicate cycle alone.

Spectroscopy of eclipse measurement and direct imaging of large super-Earths are a first steps to assess whether different chemical cycles can be detected on rocky exoplanets and their underlying conditions explored. With future missions like JWST and ground based observations we can start to explore the diversity of potentially habitable earth-like and super-Earth exoplanets. In the era of characterizing massive terrestrial planets and in view of the recent large sample of detections (Mayor et al. 2008) and the first transiting system CoRoT-7b (Leger et al. 2009), an extremely hot rocky exoplanet, the low resolution measurements proposed here can develop our understanding of geochemical cycles on rocky planets. Using the detectable features of the sulphur cycle we can assess its potential domination of the climate for dry super-Earths as one of the first steps in comparative planetology for rocky exoplanets.